\documentclass[10pt]{article}

\usepackage{amsxtra,amssymb,amsthm,amsmath,latexsym}

\usepackage{graphicx}
\newtheorem{thm}{Theorem}

\newtheorem{df}{Definition}
 \newcommand{\thmref}[1]{Theorem~\ref{#1}}

\newcommand{\bee}{\begin{equation*}}
\newcommand{\eee}{\end{equation*}}
\newcommand{\be}{\begin{equation}}
\newcommand{\ee}{\end{equation}}
\newcommand{\pn}{\par\noindent}
\title{Property $C$ and applications to inverse problems}
\author{A G Ramm\\
\small Department of Mathematics\\[-0.8ex]
\small Kansas State University, Manhattan, KS 66506-2602, USA\\[-0.8ex]
\small \texttt{ramm@math.ksu.edu}\\
}

\begin{document}
\date{}
\maketitle
\begin{abstract}Let $\ell_j:=-\frac{d^2}{dx^2}+k^2q_j(x),$
$k=const>0,\ j=1,2,$\,\, $0<c_0\leq q_j(x)\leq c_1,$
$q$ has finitely many discontinuity points $x_m\in [0,1],$ and is
real-analytic on the intervals $[x_m,x_{m+1}]$ between these points.
The set of such functions $q$ is denoted by
$M.$  Only the following property of $M$ is used: if $q_j\in M$,
$j=1,2,$ then the function
$p(x):=q_2(x)-q_1(x)$ changes sign on the interval $[0, 1]$ at most
finitely many times.
Suppose that $(*)\quad \int_0^1p(x)u_1(x,k)u_2(x,k)dx=0,\quad
\forall k>0,$ where $p\in M$ is an arbitrary  fixed function, and
$u_j$ solves the problem
$\ell_ju_j=0,\quad
0\leq x\leq 1,\quad u'_j(0,k)=0,\quad u_j(0,k)=1.$ If $(*)$ implies
$h=0$, then the pair $\{\ell_1,\ell_2\}$ is said to have property
$C$ on the set $M$. This property is proved for the pair
$\{\ell_1,\ell_2\}$.
Applications to some inverse problems for a heat equation are given.
\end{abstract}

\pn{
 }

\section{Introduction}
Property $C$ stands for completeness of the
set of products of solutions to homogeneous equations. This notion
was introduced by the author in \cite{R196}, \cite{R214},
\cite{R402}, and used widely as a powerful tool for proving
uniqueness theorems for many inverse problems
(\cite{R220}-\cite{R470}). In \cite{R402} Property C was proved
for the pair of operators $\{\frac{d^2}{dx^2}+k^2-q_1(x), 
\frac{d^2}{dx^2}+k^2-q_2(x)\}$, where $q_1, q_2 \in L_{1,1}$,
$L_{1,1}:=\{ q: q=\overline{q}, \int_0^\infty (1+x)|q(x)|dx<\infty\}$.
The novel point in our paper is the proof of Property C for the 
pair of differential operators with a different dependence on the spectral 
parameter. This new version of Property C turns to be basic, for example, 
in the proof
of uniqueness theorem for an inverse problem for a heat equation with 
a discontinuous thermal conductivity.  

The aim of this paper is to prove Property $C$ for the pair
$\{\ell_1,\ell_2\}$, where $\ell_j:=-\frac{d^2}{dx^2}+k^2q_j(x),$
$k=const>0,\ j=1,2,$ $0<c_0\leq q_j(x)\leq c_1<\infty,$ $c_0,c_1$
are constants, $q_j(x)\in M$, and $M$
is the set of real-valued integrable functions such that if $q_j\in M$,
$j=1,2,$ are arbitrary members of $M$, then the
function $p(x):=q_2(x)-q_1(x)$ changes  sign at most finitely many  
times on the interval $[0,1]$. For example, $M$ can be a set of
piecewise-analytic real-valued functions with 
finitely many discontinuity points on the interval $[0,1]$ (see 
\cite{R569}).

\begin{df}
Let $h\in M$ be an arbitrary fixed function, and
\be\label{e1}
\ell_ju_j(x,k)=0,\quad u'_j(0,k)=0,\quad u_j(0,k)=1,\quad 0\leq
x\leq 1. \ee
If the orthogonality relation
\be\label{e2}
\int_0^1h(x)u_1(x,k)u_2(x,k)dx=0\quad \forall k>0, \ee
implies $h=0$, then we say that the pair $\{\ell_1,\ell_2\}$ has Property
$C$ for the set $M$.
\end{df}

Our first result is :
\begin{thm}\label{th1}
The pair $\{\ell_1,\ell_2\}$ has Property $C$ for the set $M$.
\end{thm}
Let us give an example of applications of Property $C$. Consider the
problem:
\be\label{e3} U_t=(a(x)U')',\quad 0\leq x\leq 1,\quad t>0;\quad
U':=\frac{\partial U}{\partial x}, \ee
\be\label{e4} U(x,0)=0;\quad U(0,t)=0,\quad
U(1,t)=F(t),
\ee
\be\label{e5} a(1)U'(1,t)=G(t). \ee
Assume that
\be\label{e6} 0<c_0'\leq a(x)\leq c_1'<\infty,\quad a(x)\in M,\quad
c_0',c_1'=const. \ee
The function $F(t)\not\equiv 0,$\,
$F(t)\geq 0$, $F(t)=0$ if $t> T$, where $T>0$ is an arbitrary
fixed number, and $F\in L^1([0,T])$. Problem \eqref{e3}-\eqref{e4} has a
unique solution. The function
$G(t)$ is the measured datum (extra datum), which is the heat flux at 
the point $x=1$.
The inverse problem is:\\
$\text{IP}_1$:\,\,\, \textit{Given} $\{F(t),G(t)\}_{\forall t>0}$,
\textit{find} $a(x).$\\
The function $a(x)\in M$ may have finitely many
discontinuity points, and the solution to \eqref{e3}-\eqref{e4} is
understood in the weak sense.

Let us formulate the $IP_1$ in an equivalent but different form. Let
\be\begin{split}\label{e7} v(x,\lambda)&:=\int_0^\infty
U(x,t)e^{-\lambda t}dt,\quad f(\lambda)=\int_0^\infty F(t)e^{-\lambda
t}dt,\\ g(\lambda)&=\int_0^\infty G(t)e^{-\lambda t}dt. \end{split}\ee 
Take
the Laplace of \eqref{e3}-\eqref{e5} and get \be\label{e8}
-(a(x)v')'+\lambda v=0,\quad 0\leq x\leq 1,\quad \lambda>0, \ee
\be\label{e9} v(0,\lambda)=0,\quad v(1,\lambda)=f(\lambda),\quad
a(1)v'(1,\lambda)=g(\lambda). \ee
Let
$u(x,\lambda):=a(x)v'(x,\lambda).$ Differentiate \eqref{e8} and get
\be\label{e10} -u''+\lambda a^{-1}(x)u=0,\quad 0\leq x\leq 1,\quad
u'(0,\lambda)=0, \ee \be\label{e11} u(1,\lambda)=g(\lambda),\quad
u'(1,\lambda)=\lambda f(\lambda). \ee Let $\lambda=k^2,\quad
k>0,\quad a^{-1}(x):=q(x).$
Then
\be\label{e12}
-u''+k^2q(x)u=0,\quad 0\leq x\leq 1,\quad u'(0,k^2)=0,\quad 
u(1, k^2)=g(k^2), \ee 
and the data are $\{g(k^2),k^2f(k^2)\}_{\forall k>0}$. The IP can be
reformulated as follows:\\
$\text{IP}_2$:\,\,\, \textit{ Given the data
$\{g(k^2),k^2f(k^2)\}_{\forall k>0}$, find $q(x)$.}\\ 
Our second
result is:
\begin{thm}\label{th2}
The $\text{IP}_2$ has at most one solution.
\end{thm}

Theorem 2 implies the uniqueness of the solution to  $\text{IP}_1$
in the class of the piecewise-analytic strictly
positive functions $a(x)$ with finitely many discontinuity points on the 
interval $[0,1]$, or, more generally, in a class $M$.

In the literature $\text{IP}_1$ has been considered earlier
(see, e.g.,  \cite{R422}, \cite{R438} and references therein) in the case 
when $a(x)\in H^2([0,1])$,
where $H^2$ is the Sobolev space. For piecewise-constant 
thermal conductivity coefficients $a(x)$ with finitely many discontinuity 
points the
$\text{IP}_1$ was studied recently in \cite{R569}. An inverse problem
for equation \eqref{e3} with different extra data, namely $U(\xi_n,t)$
$\forall t>0$, $\xi_n\in [0,1]$, $1\leq n\leq N$, $\min_{1\leq n \leq N} 
|\xi_n-\xi_{n+1}|\geq \sigma >0$, where $\sigma$ is a fixed 
number, $N=3\nu$,
and $\nu$ is the number of discontinuity points of $a(x)$, was
studied in \cite{GH}. In $\text{IP}_1$ the extra data are collected at 
just
one point $x=1$.\\ 
Our arguments prove the uniqueness result for Theorem 1
in the case when the data in $\text{IP}_1$ are the values
$\{F(t),G(t)\}_{\forall t\in [0,0,T+\epsilon]}$ for an arbitrary
small $\epsilon>0$.
These data determine $a(x)$ uniquely
because the solution $U(x,t)$ is an analytic function of
$t$ in a neighborhood of the set $(T,\infty)$, so the knowledge of
$U(x,t)$ on the segment $[0,T+\epsilon]$ determines $U(x,t)$
uniquely for all $t>0$.

In Section 2 proofs are given.
\section{Proofs}
{\it Proof of \thmref{th1}.} The solution to \eqref{e1} solves
the equation \be\label{e13}
u_j(x,k)=1+k^2\int_0^x(x-s)q_j(s)u_j(s,k)ds,\quad x\geq 0,\quad
j=1,2. \ee This is a Volterra equation. It has a unique solution
$u_j(x,k)$. This solution has the following properties:
\be\label{e14} u_j(x,k)\geq 1,\quad u'_j(x,k)\geq 0,\quad
u''_j(x,k)>0,\quad 0\leq x\leq 1, \ee \be\label{e15}
\frac{\partial^i u_j}{(\partial k^2)^i}\geq 0,\quad i=1,2,3,\hdots,
\ee 
\be\label{e16} \lim_{k\to
\infty}\frac{u_j(y,k)}{u_j(x,k)}=0,\qquad 0\leq y<x\leq 1. \ee
Properties \eqref{e14}-\eqref{e15} are immediate consequences of
\eqref{e13}. Let us prove \eqref{e16}. One has \be\label{e17}
u_j(x,k)=u_j(y,k)+\int_y^x u_j'(s,k)ds. \ee 
From \eqref{e13} and \eqref{e14} one
obtains \be\label{e18} u'_j(x,k)=k^2\int_0^xq_j(s)u_j(s,k)ds\geq
k^2\int_0^xq_j(s)ds. \ee 
From equations \eqref{e17} and \eqref{e18} one gets:
\be\begin{split}
\frac{u_j(x,k)}{u_j(y,k)}&=1+\int_y^x\frac{u'_j(s,k)}{u_j(y,k)}ds=1+k^2\int_y^x\frac{\int_0^sq_j(z)u_j(z,k)dz}{u_j(y,k)}ds\\
&\geq 1+k^2\int_y^xds\int_y^sq_j(z)ds\geq
1+\frac{1}{2}k^2c_0(x-y)^2\to \infty\quad as\quad k\to \infty.
\end{split}\ee
Thus, \eqref{e16} is proved.

Since $h\in M$, the segment $[0,1]$ is a union of finitely many
intervals without common interior points on each of which
the function $h(x)$
keeps sign. Let $[z,1]$ be such an interval. We want to prove that
$h=0$ on this interval. If this is done then similarly, in a finite
number of steps, one proves that $h=0$ on the whole interval
$[0,1]$, and then the proof of 
Theorem 1 is completed. 

Let us rewrite relation
\eqref{e2} as \be\begin{split}\label{e20}
\int_z^1h(x)u_1(x,k)u_2(x,k)dx&=-\int_0^zh(x)u_1(x,k)u_2(x,k)dx\\
&\leq u_1(z,k)u_2(z,k)\int_0^z|h(x)|dx,\end{split} \ee 
where the monotonicity and the positivity of $u_j$ was used, see 
\eqref{e14}.
Without loss
of generality assume that $h(x)>0$ on $[z,1]$ and fix an arbitrary
$y\in (z,1)$. Then \be\label{e21} \int_z^1h(x)u_1(x,k)u_2(x,k)dx\geq
\int_y^1h(x)dx\,\,u_1(y,k)u_2(y,k). \ee 
From \eqref{e20} and \eqref{e21}
one gets: \be\label{e22} \int_y^1h(x)dx\leq
\frac{u_1(z,k)u_2(z,k)}{u_1(y,k)u_2(y,k)}\int_0^z|h(x)|dx,\qquad y>z.
\ee Let $k\to \infty$ in \eqref{e22} and use \eqref{e16} to get
$\int_y^1h(x)dx=0.$ Since $h(x)\geq 0$ on $[z,1]$, it follows that
$h=0$ on $[y,1]$. Since the point $y\in(z,1)$ is arbitrary, 
it follows that $h=0$ on 
$[z,1]$.\\
\thmref{th1} is proved. \hfill $\Box$\\

{\it Proof of \thmref{th2}.} Assume the contrary, i.e., there
are pairs of functions  $\{\psi_1, q_1\}$ and $\{\psi_2, q_2\}$ which 
solve \eqref{e12} and
\eqref{e11} with $\lambda=k^2$. Let $w=\psi_1-\psi_2$. Then
\be\label{e23} w'(0,k)=w(1,k)=w'(1,k)=0, \ee and \be\label{e24}
-w''+k^2q_1(x)w=k^2p(x)\psi_2,\quad p(x):=q_2(x)-q_1(x). \ee
Multiply \eqref{e24} by $u_1(x,k)$, integrate over $[0,1]$, and then
integrate by parts using \eqref{e23}. The result is: \be\label{e25}
k^2\int_0^1p(x)u_1(x,k)\psi_2(x,k)dx=0 \quad \forall k>0. \ee 
Since
$\psi_2(x)=c(k)u_2(x,k),$ where $c(k)=const\neq 0,$ and $p\in M$, it
follows from \eqref{e25} and \thmref{th1} that $p=0$, so
$q_1=q_2.$\\ \thmref{th2} is proved.  \hfill $\Box$


\begin{thebibliography}{00}
\bibitem{GH} S. Gutman, J. Ha, Identifiability of piecewise constant 
conductivity in a heat conduction process, SIAM J. Contr. Optim., 46, N2, 
(2007), 694-713.

\bibitem{R569} N.S. Hoang, A.G. Ramm, An inverse problem for a heat
equation with
piecewise-constant thermal conductivity, J. Math. Phys., 50, 063512, 
(2009).

\bibitem{R196} A.G.Ramm, On completeness of the products of
harmonic
 functions, Proc. A.M.S., 99, (1986), 253-256.

\bibitem{R214} A.G.Ramm, Completeness of the products of solutions
to PDE
 and uniqueness theorems in inverse scattering,
 Inverse problems, 3, (1987), L77-L82.

\bibitem{R220} A.G. Ramm, Multidimensional inverse problems and
completeness
 of the products of solutions to PDE. J. Math. Anal.
 Appl. 134, 1, (1988), 211-253; 139, (1989) 302.

\bibitem{R252} A.G.Ramm, Completeness of the products of solutions
of PDE
 and inverse problems, Inverse Probl.6,
 (1990), 643-664.

\bibitem{R262} A.G.Ramm, Necessary and sufficient condition for a
PDE to have property C, J. Math. Anal. Appl.156, (1991), 505-509.

\bibitem{R402} A.G.Ramm,
Property C for ODE and applications to inverse problems, in the book
"Operator Theory and Its Applications", Amer. Math. Soc., Fields
Institute Communications vol. 25, (2000), pp.15-75, Providence, RI.

\bibitem{R422} A.G.Ramm, An inverse problem for the heat equation,
Jour. of Math. Anal. Appl., 264, N2, (2001), 691-697.

\bibitem{R438} A.G.Ramm, An inverse problem for the heat equation II,
Applic. Analysis, 81, N4, (2002), 929-937.

\bibitem{R470} A.G.Ramm, {\bf Inverse problems, Springer,
New York,} 2005.
\end{thebibliography}
\end{document}